# Nonlinear optics in the electron-hole continuum in 2D semiconductors: two-photon transition, second harmonic generation and valley current injection

**Authors:** Pu Gong [1]*, Hongyi Yu[1]†, Yong Wang[2], Wang Yao [1]

**Affiliations:**

[1] Department of Physics and Center of Theoretical and Computational Physics, The University of Hong Kong, Hong Kong, China

[2] School of Physics, Nankai University, Tianjin, China

*Correspondence to: pugong.phy@gmail.com

†Correspondence to: yuhongyi@hku.hk

Abstract

**We investigate two-photon transitions to the electron-hole scattering continuum in monolayer transition-metal dichalcogenides, and identify two contributions to this nonlinear optical process with opposite circularly polarized valley selection rules. In the non-interacting limit, the competition between the two contributions leads to a crossover of the selection rule with the increase of the two-photon energy. With the strong Coulomb interaction between the electron and hole, the two contributions excite electron-hole scattering states in orthogonal angular momentum channels, while the strength of the transition can be substantially enhanced by the interaction. Based on this picture of the two-photon transition, the second harmonic generation (SHG) in the electron-hole continuum is analyzed, where the Coulomb interaction is shown to greatly alter the relative strength of different cross-circular polarized SHG processes. Valley current injection by the quantum interference of one-photon and two-photon transition is also investigated in the presence of the strong Coulomb interaction, which significantly enhances the injection rate.**

**Main Text**

**1. Introduction**

Monolayer transition-metal dichalcogenides (TMDs) have emerged as a new class of semiconductors in the atomically thin two-dimensional limit [1-4]. They feature a direct band gap in the visible frequency range, with the band edges located at the degenerate but inequivalent K and -K corners of the hexagonal Brillouin zone [5-7].

The degenerate band edges at K and -K are characterized by an extra quantum degree of freedom known as valley. The low energy electrons and holes are described by a massive Dirac Fermion model, where the interband optical transitions feature a valley-dependent selection rules that allows the selective excitation of K and -K valleys by absorbing a circular polarized photon [8-10]. Generation and detection of valley polarization and coherence can therefore be realized through the light-matter interaction [11-14], allowing access to various valley related phenomena using linear optical processes [15-20]. With the exceptionally strong Coulomb interaction in monolayer TMDs because of its 2D geometry and the large effective mass, a pair of optically excited electron and hole can form tightly bound hydrogen-like state known as exciton [21-28]. The exciton Bohr radius is in the order of a few nm, still much larger compared to the lattice constant. Therefore, in the momentum space, the exciton wavefunction is well localized at the K and -K valleys, which inherit the valley optical selection rules of the interband transitions. The various excitonic ground and excited states at the K and -K valleys can then be selectively accessed through the one-photon process [29].

Two-photon transition is a fundamental nonlinear optical process which follows different selection rules from the one-photon process [30-33]. Monolayer TMDs as a direct-gap semiconductor provides a new platform for nonlinear optical studies in the ultimate 2D limit [34]. Two-photon measurements in monolayer TMDs have been implemented to measure the resonances of excitonic excited states that are dark in one-photon process, for a determination of the exciton Rydberg series and the detail of the

Coulomb binding [35-39]. For the excitonic bound states, selection rules for the two-photon process can be similarly determined from rotational symmetry [40-42], just like the one-photon optical selection rules. Second-harmonic generation (SHG) is another nonlinear optical process of fundamental importance, which is possible in materials lacking of inversion symmetry [43-44]. Second-harmonic signals have been used for revealing the symmetry and crystalline orientation of 2D TMDs [45-53]. It has been shown that excitonic bound states can significantly enhance the SHG when they are in resonance with the two-photon energy [38-39, 54-56]. Based on such resonance enhancement, remarkable electrical tunability of SHG has been demonstrated in monolayer TMD transistor [57].

While most existing studies have focused on the excitonic bound states, nonlinear optical transitions into the electron-hole continuum (i.e. the continuum of scattering states with energies above the single-particle band) are also of interest. A prominent example is the optical injection of spin and charge currents through the interference of two-photon and one-photon transition in the electron-hole continuum, which been extensively studied in bulk and quantum well structures of semiconductors [58-64]. The scheme has been experimentally implemented in generating ballistic charge current in 2D materials as well [65-66]. In a non-interacting picture, the quantum interference of one- and two-photon excitation by two light beams lead to unbalanced interband transitions at a pair of momentum space points of opposite band velocities, which then correspond to a current of the photo-carriers that can be controlled by the relative phases of the two light fields [58]. Theoretical studies in the non-interacting limit have also generalized this two-color quantum interference scheme for valley current injection in monolayer TMDs [67]. With the strong Coulomb interaction in monolayer TMDs, however, the one-photon and the two-photon transition cannot be described by interband transitions at individual momentum space points. The continuum of scattering states under the strong Coulomb interaction between the electron and hole need to be used instead. The excitonic effect in the continuum can significantly change the current generation efficiency, as well as the physical description of the current injection [63].

In this paper, we investigate two-photon nonlinear optical transitions to the electron-hole continuum in monolayer TMDs, and identify two contributions with opposite circularly polarized valley optical selection rules. The first contribution consists of two interband transitions mediated by a remote band. It has a valley optical selection rule opposite to that of the one-photon transition, and the magnitude is insensitive to the two-photon energy. The second is a contribution within the massive Dirac cone, consisting of an interband transition and an intraband one, and the selection rule is the same as the one-photon transition. In the non-interacting limit, the second contribution is proportional to the band velocity which vanishes at the band edge and grows fast with the increase in two-photon energy, and consequently there is a crossover in strength between the two contributions, accompanied by a reversal in the valley optical selection rule. Going beyond the non-interacting picture, the strong Coulomb interaction between electron and hole is taken into account through a 2D hydrogen model with modified Bohr radius for describing the electronic states involved in the optical transition. The two contributions excite electron-hole scattering states in orthogonal channels respectively, and we find the Coulomb interaction can substantially enhance the two-photon transition strength from the second contribution (within the massive Dirac cone), which becomes finite at the edge of the electron-hole continuum. Based on these results on two-photon transition matrix elements, the SHG in the electron-hole continuum is analyzed, and Coulomb interaction is shown to greatly alter the relative strength of different cross-circular polarized SHG process. Moreover, the quantum interference of one-photon and two-photon transition is also investigated in the presence of the strong Coulomb interaction, where the interference of the s-wave and p-wave electron-hole scattering states excited respectively by light fields of the two colors lead to the ballistic current injection. Under the co-circular polarized two-color excitations, we show the valley current injection controlled by the polarization, where Coulomb interaction significantly enhances the injection rate.

## 2. Non-interacting limit: polarization reversal in two-photon transition

We consider two-photon transition from the valance band to conduction band. We begin with the non-interacting case in which the Coulomb interaction of electrons is neglected. The two-photon transition rate is given by

$$W = \frac{2\pi}{\hbar} \sum_{\mathbf{k}} |M(\mathbf{k})|^2 \delta(E_c(\mathbf{k}) - E_v(\mathbf{k}) - 2\hbar\omega), \qquad (1)$$

where the two-photon matrix element is

$$M(\mathbf{k}) = \frac{e^2 E_\omega^2}{m_0^2 \omega^2} \sum_i \frac{\langle f|\boldsymbol{\epsilon} \cdot \hat{\mathbf{p}}|i\rangle \langle i|\boldsymbol{\epsilon} \cdot \hat{\mathbf{p}}|0\rangle}{E_i(\mathbf{k}) - E_v(\mathbf{k}) - \hbar\omega}, \qquad (2)$$

where $\mathbf{E}_\omega = E_\omega e^{-i\omega t} \boldsymbol{\epsilon} + \text{c.c.}$ is the light field in frequency $\omega$ with $\boldsymbol{\epsilon}$ the polarization vector of the field. The initial state is the ground state $|0\rangle$ with completely filled valence band and empty conduction band. In non-interacting case, the final state $|f\rangle = a_{c,\mathbf{k}}^\dagger a_{v,\mathbf{k}} |0\rangle$ is a free electron-hole pair state formed by conduction ($c$) and valence ($v$) bands at the wave vector $\mathbf{k} = k(\cos\theta, \sin\theta)$, which is the relative wave vector measured from $\pm K$ point. $e$ and $m_0$ are the free electron charge and mass. The sum of intermediate state $|i(\mathbf{k})\rangle$ runs over different electronic band indexes $i = c, v, \ldots$ and $E_i(\mathbf{k})$ the band energy at $\mathbf{k}$. As valley index is conserved in optical transitions, here we focus on the transitions in one (K) valley, and the results of $-K$ valley can be obtained from the time reversal operation.

In monolayer TMDs, the material displayed valley-dependent optical selection rules for $\sigma^\pm$ (right- and left-handed) circular polarized light [9], at K point the one-photon interband transition is excited by $\sigma^\pm$ light. For two-photon transition, we also consider the excitation of $\sigma^+$ light denoted by the polarization vector $\boldsymbol{\epsilon}_\pm = \hat{\mathbf{x}} \pm i\hat{\mathbf{y}}$. The two-photon excitation frequency satisfies $2\hbar\omega \geq E_g$ where $E_g$ is the band gap, in calculations of this paper we take $E_g = 2.5\text{eV}$ which is at typical range of band gap for monolayer TMDs. We divide the two-photon matrix element into two parts according to the contributions from different intermediate band states, $M^\pm(\mathbf{k}) = M_r^\pm(\mathbf{k}) + M_{cv}^\pm(\mathbf{k})$, where

$$M_r^\pm(\mathbf{k}) = \frac{e^2 E_\omega^2}{m_0^2 \omega^2} \sum_{i \neq c,v} \frac{p_{ci}^\pm(\mathbf{k}) p_{iv}^\pm(\mathbf{k})}{E_i(\mathbf{k}) - E_v(\mathbf{k}) - \hbar\omega}, \qquad (3a)$$

$$M_{cv}^\pm(\mathbf{k}) = \frac{e^2 E_\omega^2}{m_0^2 \omega^2} \left[ \frac{p_{cv}^\pm(\mathbf{k})\left(p_{cc}^\pm(\mathbf{k}) - p_{vv}^\pm(\mathbf{k})\right)}{\hbar\omega} \right]. \qquad (3b)$$

Here $M_{cv}^\pm(\mathbf{k})$ is the contribution by intermediate sates from the conduction and valence bands, $M_r^\pm(\mathbf{k})$ represents the contribution from the bands other than conduction and valence bands referred as remote ($r$) bands. Here $p_{ij}^\pm(\mathbf{k}) = p_{ij}^x(\mathbf{k}) \pm ip_{ij}^y(\mathbf{k})$, with $p_{ij}^\alpha(\mathbf{k}) = \left\langle u_{i,\mathbf{k}} \left| \frac{m_0}{\hbar} \frac{\partial H_0}{\partial k_\alpha} \right| u_{j,\mathbf{k}} \right\rangle$ the interband (intraband) momentum matrix element near $\pm K$ point for $i \neq j$ ($i = j$), $H_0$ is the single-particle Hamiltonian of the monolayer and $u_{i,\mathbf{k}}$ is the Bloch periodic part of band state. The momentum matrix elements $p_{ij}^\pm(\mathbf{k})$ can be calculated by perturbation expansion [29]. Due to the discrete three-fold rotation symmetry of the material, at $\pm K$ point ($k = 0$) the electronic state for band $i$ has a three-fold rotation quantum number $m_i$. For band indexes $i, j$ there is $\Delta m_{ij} \equiv m_i - m_j = 3N, \mp 1 + 3N$ where $N$ is an integer [3]. If $\Delta m_{ij} = \mp 1 + 3N$, the interband one-photon transition by $\sigma^\pm$ ($\sigma^\mp$) excitation will be dipole-allowed (dipole-forbidden), which refers to finite (zero) transition strength at the band extrema K point. The corresponding interband momentum matrix elements $p_{ij}^\pm(\mathbf{k})$ ($p_{ij}^\mp(\mathbf{k})$) in K valley [29] are

$$p_{ij}^\pm(\mathbf{k}) = \alpha_{ij} + \beta_{ij} k^2 + O(k^3), \qquad (4a)$$

$$p_{ij}^\mp(\mathbf{k}) = \gamma_{ij} k e^{\pm i\theta_\mathbf{k}} + \xi_{ij} k^2 e^{\mp 2i\theta_\mathbf{k}} + O(k^3), \qquad (4b)$$

where $\alpha_{ij}$, $\beta_{ij}$, $\gamma_{ij}$, $\xi_{ij}$ are the parameters depending on band combinations. $\theta_\mathbf{k}$ is the polar angle at $\mathbf{k}$ space. For $\Delta m_{i'j'} = 3N$, with $\sigma^\pm$ one-photon excitation we have

$$p_{i'j'}^\pm(\mathbf{k}) = b_{i'j'} k e^{\pm i\theta_\mathbf{k}} + c_{i'j'} k^2 e^{\mp 2i\theta_\mathbf{k}} + O(k^3), \qquad (5)$$

as in such case at K point the optical transition is forbidden with $p_{i'j'}^\pm(0) = 0$, only at

$k \neq 0$ where the three-fold rotation is not conserved that $p^{\pm}_{i'j'}(\mathbf{k})$ is finite. Specifically, for the intraband momentum matrix elements for $c$ and $v$ bands $p^{\pm}_{cc}(\mathbf{k})$, $p^{\pm}_{vv}(\mathbf{k})$ appearing at $M^{\pm}_{cv}(\mathbf{k})$, there is $p^{\pm}_{cc}(\mathbf{k}) - p^{\pm}_{vv}(\mathbf{k}) = \frac{\hbar}{\mu_{cv}} k e^{\pm i\theta_{\mathbf{k}}} + O(k^2)$ with the reduced mass $\mu_{cv}$ defined as $1/\mu_{cv} = 1/m_c + 1/m_v$ by the convention that the conduction (valence) band electron mass $m_{c(v)} > 0$.

| band indexes | | $p^+_{ij}(0)/p^+_{cv}(0)$ | $p^-_{ij}(0)/p^+_{cv}(0)$ |
|---|---|---|---|
| $i$ | $j$ | | |
| $c$ | $v$ | 1 | 0 |
| $c$ | $v-4$ | 0.42 | 0 |
| $c$ | $v-3$ | 0 | 0.82 |
| $c+2$ | $v$ | 0 | 0.30 |
| $v-3$ | $v$ | 0 | 0.20 |
| $v-5$ | $v$ | 0.085 | 0 |

**Table I**: $p^+_{ij}$ values at K point from the first-principle calculations. For up to 11 bands where there are 4 upper bands ($c, c+1, \ldots c+3$) above the band gap and 7 lower bands ($v, v-1, \ldots v-6$) below the gap, the $p^+_{ij}(0)$ values with non-negligible contributions $> 1 \times 10^{-3} p^+_{cv}(0)$ are listed.

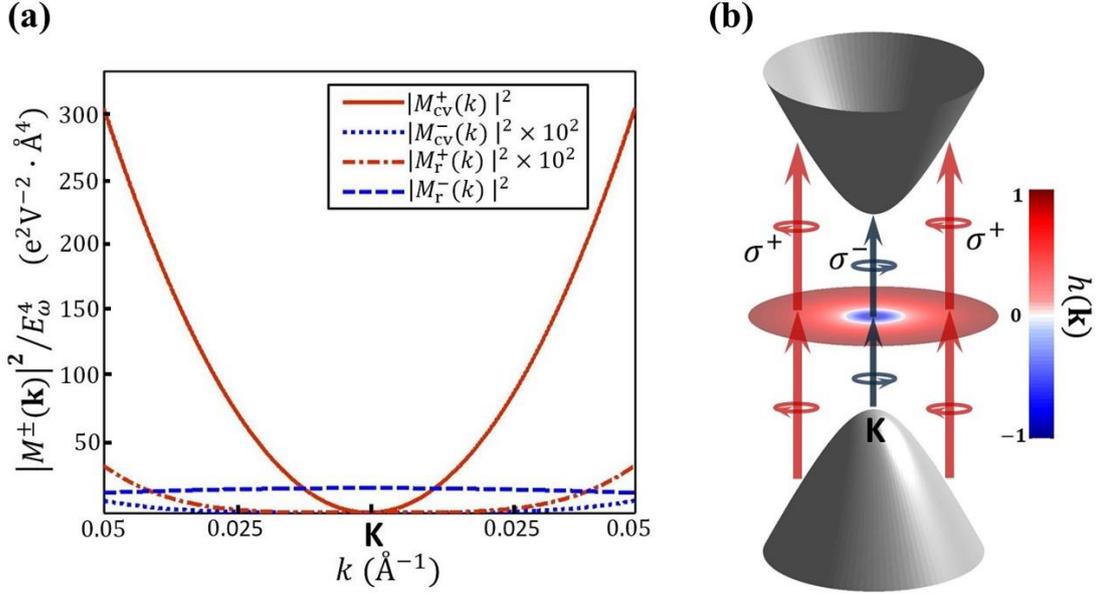

**Figure 1**: **(a)** Strengths of two-photon transition matrix elements at K valley with circular-polarized excitation. The matrix elements are divided into two contributions according to the band indexes in intermediate states, in which $M_{cv}^{\pm}(\mathbf{k})$ is the contribution from c and v bands and $M_{r}^{\pm}(\mathbf{k})$ is by remote ($r$) bands respectively. **(b)** Polarization reversal of two-photon transition at K valley. The colored disk illustrated is the degree of two-photon circular polarization $h(\mathbf{k})$ which reversed from $-1$ at K point to $1$ at larger $k$.

The values of $p_{ij}^{\pm}(\mathbf{k})$ are obtained by fitting first-principle calculation results for monolayer WS2. In **Table I** the values at K point $p_{ij}^{\pm}(0)$ with innegligible contributions are listed for different band index combinations. Behavior of $|M_{cv}^{\pm}(\mathbf{k})|^2$ and $|M_{r}^{\pm}(\mathbf{k})|^2$ is displayed in **Fig. 1(a)**. At K point, as the intraband momentum matrix elements is zero, the two-photon $\sigma^{\pm}$ transition contributed by $c$ and $v$ bands vanishes, and $M_{cv}^{\pm}(\mathbf{k})$ increases as k becomes finite and larger. While for the contribution from the $r$ bands, $M_{r}^{-}(\mathbf{k})$ is non-zero at K point, and the magnitude $|M_{r}^{-}(\mathbf{k})|$ is merely dependent on $k$. The behavior of $|M_{cv}^{+}(\mathbf{k})|$ and $|M_{r}^{-}(\mathbf{k})|$ is consistent with the discrete three-fold rotation symmetry of the material.

At K point the one-photon-allowed excitation is by $\sigma^+$ polarized light, the two-photon transition is by $\sigma^-$ light, as $\Delta m_{cv} + m_{ph} = 3N$ for $m_{ph} = -2$ (+1) for two $\sigma^-$ (one $\sigma^+$) photons. Away from K point, with the increase of $k$ the two-photon transition displays reversal of circular polarization [see **Fig. 1 (a)**]. At smaller $k$, the two-photon transition is dominated by two $\sigma^-$ process, as $k$ increases the circular polarization of two-photon transition reverses to a two $\sigma^+$ transition due to the increase of $|M_{cv}^+(\mathbf{k})|$. We illustrated such polarization reversal by the degree of two-photon polarization $h(\mathbf{k}) = [|M^+(\mathbf{k})|^2 - |M^-(\mathbf{k})|^2]/[|M^+(\mathbf{k})|^2 + |M^-(\mathbf{k})|^2]$ in **Fig. 1(b)** as $M^\pm(\mathbf{k}) = M_{cv}^\pm(\mathbf{k}) + M_r^\pm(\mathbf{k})$. At K point, $h(0) = -1$ and as the increase of $k$, $h(\mathbf{k})$ changes from $-1$ to $h(\mathbf{k}) \sim 1$ at larger $k$, the two-photon polarization changing from $\sigma^-$ to $\sigma^+$.

## 3. Two-photon transition to electron-hole scattering continuum under strong Coulomb interaction

In the previous section we studied two-photon transition without electron-hole Coulomb interaction. While such interaction has been demonstrated to play important role in 2D TMDs, in this section we include electron-hole interaction to the problem. With the interaction the electron-hole pair states at different $\mathbf{k}$ as $a_{c,\mathbf{k}}^\dagger a_{v,\mathbf{k}}|0\rangle$ are coupled, the state of interest becomes $|cvX\rangle = \sum_{\mathbf{k}} \varphi_X(\mathbf{k}) a_{c,\mathbf{k}}^\dagger a_{v,\mathbf{k}}|0\rangle$ where $\varphi_X(\mathbf{k})$ is the envelope function for the relative motion of electron and hole. The matrix element of two-photon transition is

$$M_X = \frac{e^2 E_\omega^2}{m_0^2 \omega^2} \sum_{c'v'X'} \frac{\langle cvX|\boldsymbol{\epsilon}\cdot\hat{\mathbf{p}}|c'v'X'\rangle \langle c'v'X'|\boldsymbol{\epsilon}\cdot\hat{\mathbf{p}}|0\rangle}{E_{c'v'}^{X'} - \hbar\omega}, \qquad (6)$$

The intermediate states $|c'v'X'\rangle$ with energy $E_{c'v'}^{X'}$ are the states with attractive Coulomb interaction by $c'$ band electron and $v'$ band hole. The sum of intermediate states runs for all the band indexes with all possible configurations of $\varphi_{X'}$. For the envelope function $\varphi_X$, we take the 2D hydrogen model [68] as a tentative description.

There are two kinds of solutions of the model: bound states known as excitons in discrete energy levels located within the band gap, and scattering states above the band edge which are also known as electron-hole continuum states. For the two-photon transition above the band edge, the scattering electron-hole continuum solution shall be the final state of interest. In the 2D hydrogen wavefunction the scattering continuum can be denoted as $\psi_{cv}^{\eta,m}(\mathbf{r}) = \sum_\mathbf{k} \varphi_{cv}^{\eta,m}(\mathbf{k}) e^{i\mathbf{k}\cdot\mathbf{r}}$[68, 69], with the energy $E_{cv}^\eta = \frac{\hbar^2 \eta^2}{2\mu_{cv}} + E_{cv}^g$ where $\eta$ is the wave number for continuum state and m = $0, \pm 1, \pm 2 \ldots$ is the angular momentum quantum number for $s, p, d, \ldots$ scattering waves. Such state is denoted by $X \equiv (\eta, m)$.

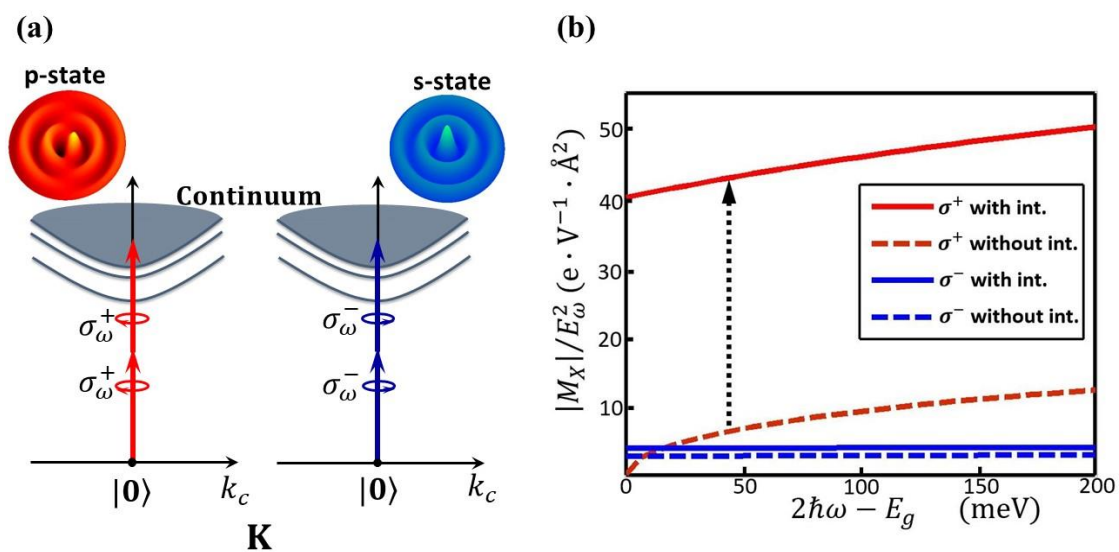

**Figure 2**: **(a)** Selection rules for two-photon transition to electron-hole continuum. Two-photon transition for $\sigma^+$ ($\sigma^-$) light excites the p-state (s-state). (b) Magnitudes of two-photon matrix elements $|M_X|$ with/without the attractive Coulomb interaction. The two $\sigma^+$ process (red solid line) is stronger than the two $\sigma^-$ one (blue solid line). In the case without interaction, the polarization reversal behavior discussed in last section is displayed. Giant enhancement of two-photon transition strength in the $\sigma^+$ process is illustrated (denoted by the dash up-arrow).

In our previous work, excitons in monolayer TMDs are shown to possess discrete three-fold rotation invariance [29], allowing s-state exciton to be one-photon bright with circularly polarized light. It has been demonstrated in recent works that the excitons which are two-photon bright are p-states [36-39]. The electron-hole scattering continuum states possess the same symmetry as the bound excitons, we consider the two-photon transition to such continuum states with circularly polarized light. At K valley the two-photon matrix elements of $\sigma^+$ ($\sigma^-$) polarized excitation are (see also supplementary information)

$$M_{\eta,1}^{(2),+} = \frac{-e^2 E_\omega^2 \sqrt{S}}{m_0^2 \omega^2} \sum_{c'v'} (\delta_{v,v'} b_{c,c'} - \delta_{c,c'} b_{v,v'}) p_{c'v'}^+(0)$$
$$\times \int [\psi_{cv}^{\eta,1}(\mathbf{r})]^* (\boldsymbol{\epsilon}_+ \cdot \nabla_\mathbf{r}) G_{c'v'}(\mathbf{r}, 0, E_{c'v'}^g/\hbar - \omega) d^2r,$$
(7a)

$$M_{\eta,0}^{(2),-} = \frac{-e^2 E_\omega^2 \sqrt{S}}{m_0^2 \omega^2} \sum_{c'v'} \left(\delta_{v,v'} p_{cc'}^-(0) - \delta_{c,c'} p_{vv'}^-(0)\right) p_{c'v'}^-(0)$$
$$\times \int [\psi_{cv}^{\eta,0}(\mathbf{r})]^* G_{c'v'}(\mathbf{r}, 0, E_{c'v'}^g/\hbar - \omega) d^2r,$$
(7b)

Here $M_{\eta,1}^{(2),+}$ ($M_{\eta,0}^{(2),-}$) denotes matrix elements for the two $\sigma^+$ ($\sigma^-$) transitions at K valley with the selection rules to p-state (s-state) with m = 1 (m = 0) and $S$ is the area of the material. These two terms are consistent with the symmetry analysis on the optical selection rules of excitonic states with rotational symmetry [29, 38-39]. Such two orthogonal angular momentum channels of two-photon transitions for circularly polarized light play the key role in our paper. The channel of p-state by $\sigma^+$ light $M_{\eta,1}^{(2),+}$ is mainly from the contribution of $c$ and $v$ bands, while the channel of s-state by $\sigma^-$ light $M_{\eta,0}^{(2),-}$ is mainly contributed by $r$ bands. Here $G_{c'v'}(\mathbf{r}, \mathbf{r}', \omega) = \sum_X [\psi_{c'v'}^X(\mathbf{r}) \psi_{c'v'}^{X*}(\mathbf{r}')]/(E_{c'v'}^X - \hbar\omega)$ is the Coulomb Green function [70] including all the bound and scattering intermediate states. In Fig. 2(b) we demonstrated the two-photon transition matrix elements for $\sigma^+$ ($\sigma^-$) light of frequency $\omega$ with excess two-photon energy $2\hbar\omega - E_g \geq 0$ satisfying $2\hbar\omega \sim E_{cv}^\eta > E_g$. For monolayer TMDs we take $a_{cv} = 1.5$nm, the value of Bohr radius from first-principle calculations on WS$_2$ [36]. We compared the matrix elements with electron-hole interaction to the non-interacting case. In **Fig. 2 (b)**, by including the interaction, the strength of two $\sigma^+$ process is greatly enhanced and became the dominant process for two-photon transition of electron-hole continuum. The polarization reversal discussed in non-interacting case indicated by the intersect of the two dash curves in **Fig. 2 (b)**, is "submerged" by the two $\sigma^+$ process. The strength of two $\sigma^-$ process contributed by the s-state, which is originated from the remote band contributions, does not change much after including attractive interaction.

We can understand the enhancement in two-photon transition for $\sigma^+$ channel from two aspects. (1) The final state is the electron-hole continuum $|cvX\rangle = \sum_{\mathbf{k}} \varphi_X(\mathbf{k}) a_{c,\mathbf{k}}^\dagger a_{v,\mathbf{k}} |0\rangle$ whose two-photon transition strength is the weighted average of the $\mathbf{k}$-point two-photon matrix element $M_X(\mathbf{k})$. In the non-interacting case the summation only includes those $\mathbf{k}$ with the magnitudes $k \propto \sqrt{2\hbar\omega - E_g}$. The $\sigma^+$ polarized two-photon absorption of these states is zero at $k = 0$ and increases with the increasing of $k$ (see the solid red line in **Fig. 1(a)**), while the $\sigma^-$ polarized two-photon absorption is nearly a constant (dashed blue line in **Fig. 1(a)**). When the Coulomb interaction is included, states with $k > 0$ are mixed in $|cvX\rangle$, which significantly contributes to the $\sigma^+$ polarized two-photon absorption but not to the $\sigma^-$ polarized case. (2) The two-photon transition matrix element $M_X$ involves the factors $\frac{1}{E_{c'v'}^{X'} - \hbar\omega}$ with $E_{c'v'}^{X'}$ the intermediate state energy [**Equation (6)**]. Intermediate states of $\sigma^+$ channel include the bound exciton states formed between $c$ and $v$ bands. Due to the strong Coulomb effects of such exciton states, the energies of such excitons $E_{cv}^X$ are relatively closer to the one-photon energy $\hbar\omega$, giving rise to larger contributions to the factor $\frac{1}{E_{cv}^X - \hbar\omega}$ compared to the non-interacting case ($\leq \frac{1}{E_g - \hbar\omega}$). While for the $\sigma^-$ channel, as the contributed bands are remote bands, due to the much larger mismatch of remote band exciton energies with one-photon energy, inclusion of Coulomb effects for remote bands have limited modification to the factor $\frac{1}{E^{X'}_{c'v'}-\hbar\omega}$, hence limited enhancement in $\sigma^-$ channel.

## 4. Second-harmonic generation above the band edge

After the two-photon transition, we turn to the second-harmonic generation (SHG) above the band edge from electron-hole continuum states. The matrix element describing SHG is

$$S_X = \frac{eE^*_{2\omega}}{2m_0\omega} \frac{\langle 0|(\boldsymbol{\epsilon}_{2\omega}\cdot\hat{\mathbf{p}})^\dagger|cvX\rangle}{(E^X_{cv} - 2\hbar\omega + i\Gamma_X)} M_X, \tag{8}$$

where $\Gamma_X$ is the decay rate of the emission state with $E^X_{cv} \sim 2\hbar\omega > E_g$ and $E^*_{2\omega}$ is the complex conjugate amplitude of light field of frequency $2\omega$. The SHG is a two-photon excitation to $|cvX\rangle$ state characterized by $M_X$ in **Equation (6)** followed by a one-photon emission from the same state. For circularly polarized excitation, symmetry analysis [71] indicates that the SHG is a cross-circular polarized process for three-fold rotation invariant materials. Hence the SHG processes of interest are $\sigma^+_\omega + \sigma^+_\omega \to \sigma^-_{2\omega}$ and $\sigma^-_\omega + \sigma^-_\omega \to \sigma^+_{2\omega}$ given by (see also supplementary information)

$$S^{++,-}_{\eta,1} = \frac{\left[M^{(1),-}_{\eta,1}\right]^* M^{(2),+}_{\eta,1}}{\left(E^\eta_{cv} - 2\hbar\omega + i\Gamma_X\right)}, \tag{9a}$$

$$S^{--,+}_{\eta,0} = \frac{\left[M^{(1),+}_{\eta,0}\right]^* M^{(2),-}_{\eta,0}}{\left(E^\eta_{cv} - 2\hbar\omega + i\Gamma_X\right)}, \tag{9b}$$

$M^{(2),+}_{\eta,1}$, $M^{(2),-}_{\eta,0}$ are two-photon transition matrix elements to p-state (s-state) by $\sigma^+$ ($\sigma^-$) light discussed in last section. The two orthogonal angular momentum channels for p- and s-state transitions correspond to the $\sigma^+_\omega + \sigma^+_\omega \to \sigma^-_{2\omega}$ and $\sigma^-_\omega + \sigma^-_\omega \to \sigma^+_{2\omega}$ SHG process respectively. The one-photon emission matrix elements in SHG are given by $M^{(1),-*}_{\eta,1} = \frac{-ieE^*_{2\omega}\sqrt{S}}{2m_0\omega}\langle 0|(\boldsymbol{\epsilon}_-\cdot\hat{\mathbf{p}})^\dagger|cv,\eta,1\rangle$ and $M^{(1),+*}_{\eta,0} =$

$\frac{-ieE_{2\omega}^*\sqrt{S}}{2m_0\omega}\langle 0|(\epsilon_+ \cdot \hat{\mathbf{p}})^\dagger|cv,\eta,0\rangle$ for p-state and s-state respectively, with $\left|M_{\eta,0}^{(1),+}\right| > \left|M_{\eta,1}^{(1),-}\right|$. We assume the decay rates of p- and s-state formed by the $c$ and $v$ band electron-hole pairs are both $\Gamma$ as the states are energetically degenerate. In **Fig. 3(a)** we plot the strength of matrix elements for the two cross-polarized cases. Without the electron-hole interaction, the process $\sigma_\omega^- + \sigma_\omega^- \to \sigma_{2\omega}^+$ is qualitatively stronger. With the interaction, the process $\sigma_\omega^+ + \sigma_\omega^+ \to \sigma_{2\omega}^-$ is enhanced and becomes a larger one due to the increase in $M_{\eta,1}^{(2),+}$ for two $\sigma^+$ photon excitation. The attractive Coulomb interaction qualitatively alters the relative strength of the SHG process of the two orthogonal channels in electron-hole continuum regime. The two-photon excitation processes for such two cross-circular SHG channels are the two-photon channels as we discussed in above section. As the frequency doubling emission matrix elements are with $\left|M_{\eta,0}^{(1),+}\right| \sim \left|M_{\eta,1}^{(1),-}\right|$, together with the behavior of $M_{\eta,1}^{(2),+}$ and $M_{\eta,0}^{(2),-}$, the relative strength of the two cross-circular SHG channels $\sigma_\omega^+ + \sigma_\omega^+ \to \sigma_{2\omega}^-$ ($\sigma_\omega^- + \sigma_\omega^- \to \sigma_{2\omega}^+$) will be qualitatively as in two-photon $\sigma^+$ ($\sigma^-$) channels.

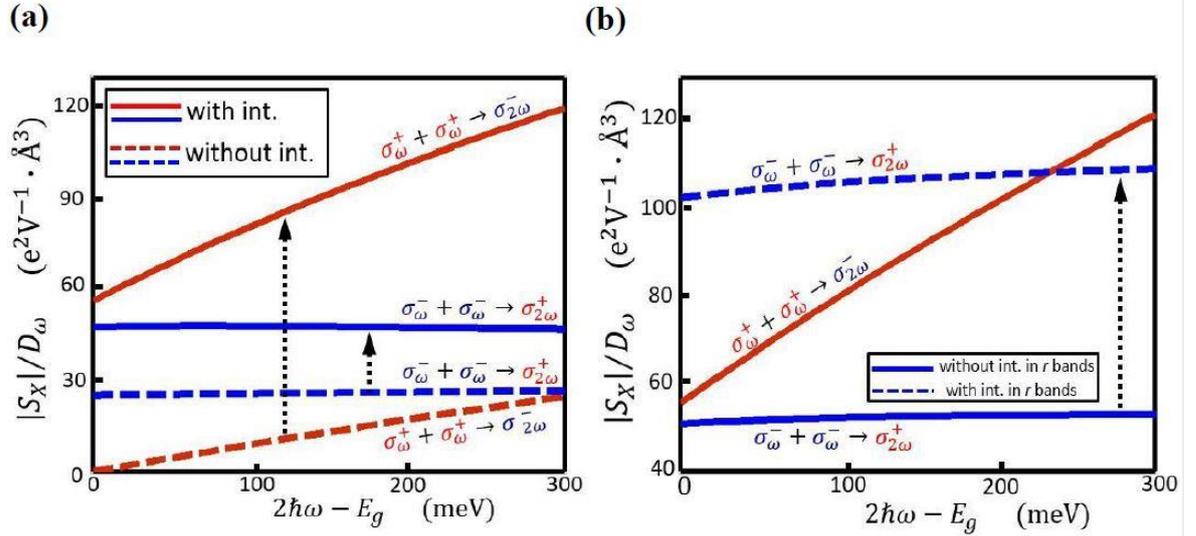

**Figure 3**: **(a)** Magnitudes of SHG elements from the excitation energy $2\hbar\omega \geq E_g$ at K valley. $\sigma_\omega^+ + \sigma_\omega^+ \to \sigma_{2\omega}^-$ process with the electron-hole Coulomb interaction (red solid line) is stronger than the $\sigma_\omega^- + \sigma_\omega^- \to \sigma_{2\omega}^+$ process (blue solid line). The non-interacting cases (red and blue dash lines) are also showed for comparison.

Enhancement by the electron-hole interaction is denoted (dash up-arrows) in both processes. (b) Coulomb interaction effects for SHG from remote-band intermediate states. The interaction in remote bands mainly influences the $\sigma_\omega^- + \sigma_\omega^- \to \sigma_{2\omega}^+$ process. The case with the interaction (blue dash line) of $a_{c'v'} \to a_0$ displays a stronger strength (the dash up-arrow) when compared with the case without interaction $a_{c'v'} \to \infty$(blue solid line). In the figure $D_\omega = E_{2\omega}E_\omega^2/\Gamma$.

As discussed in last section, the two photon $\sigma^-$ transition is contributed by the remote band states, such band states also contribute to $\sigma_\omega^- + \sigma_\omega^- \to \sigma_{2\omega}^+$ SHG process. We can discuss the influence of the remote bands' electron-hole interaction to SHG. We vary the Bohr radius $a_{c'v'}$ of the remote band excitonic states to simulate different strengths of remote band electron-hole interaction. We set $a_{c'v'} \to a_0$ for strong interaction limit, as the lower bound of $a_{c'v'}$ shall not be smaller than the lattice constant $a_0$, while $a_{c'v'} \to \infty$ is for free electron-hole pair limit of remote bands. In **Fig. 3(b)**, for free remote band electron-hole pairs, $\sigma_\omega^+ + \sigma_\omega^+ \to \sigma_{2\omega}^-$ is stronger, while with strong remote band Coulomb effects, $\sigma_\omega^- + \sigma_\omega^- \to \sigma_{2\omega}^+$ process is greatly enhanced. Such analysis reveals that, unlike the two-photon $\sigma^-$ transition in which the electron-hole interaction of remote bands has limited influence, the Coulomb interaction of remote band electron and hole may qualitatively change the SHG process in continuum regime.

Compared with **Fig. 3(a)** in which the significant enhancement of the SHG process $\sigma_\omega^+ + \sigma_\omega^+ \to \sigma_{2\omega}^-$ comes from the increased two-photon absorption of $\sigma_\omega^+ + \sigma_\omega^+$, in **Fig. 3(b)** the two-photon absorption $\sigma_\omega^- + \sigma_\omega^-$ mainly involves the remote band intermediate state. Thus an increase to the remote band Coulomb interaction can enhance the SHG process $\sigma_\omega^- + \sigma_\omega^- \to \sigma_{2\omega}^+$. On the other hand, the remote band Coulomb interaction barely affects the two-photon absorption of $\sigma_\omega^+ + \sigma_\omega^+$, so the SHG process $\sigma_\omega^+ + \sigma_\omega^+ \to \sigma_{2\omega}^-$ is not affected.

**5. Valley current injection by one- and two-photon interference**

In this section we consider applying one- and two-photon transitions together for valley current injection. For the two-color light field $\mathbf{E}(t) = E_{2\omega} e^{i(\varphi_{2\omega} - 2\omega t)} \boldsymbol{\epsilon}_{2\omega} + E_{\omega} e^{i(\varphi_{\omega} - \omega t)} \boldsymbol{\epsilon}_{\omega} + $ c.c. with $2\hbar\omega > E_g > \hbar\omega$, the field induces one-photon transition with frequency $2\omega$ and two-photon transition with $\omega$ both at the same energy above the band gap. In non-interacting case, the quantum interference of the two-color field creates carrier population imbalance $\mathbf{k}$ and $-\mathbf{k}$ [**Fig. 4(a)**], with different polarization combinations of the two fields, generation of various types of currents were demonstrated in literature [58-64]. However, in monolayer TMDs the Coulomb interaction of electron and hole is significant, which leads to pronounced effects in nonlinear optical process as we displayed in previous sections. We shall consider the interacting case for two-color quantum interference in electron-hole scattering continuum.

Here we consider the excitation scheme of two co-circularly polarized $\sigma^+$ beams at K valley, by including the electron-hole interaction the two-photon $\sigma^+$ process will excite the p-state channel in the continuum while the one-photon process will excite s-state channel [**Fig. 4(b)**] [63]. In the continuum the excited p- and s-state are energetically degenerate, the instantaneous excited state $f_\eta^{+,+}(\mathbf{r})$ under the two-color excitation is a superposition of $s$-state and $p$-state given by

$$f_\eta^{+,+}(\mathbf{r}) = \frac{M_{\eta,0}^{(1),+}}{N_\eta^+} \psi_{cv}^{\eta,0}(\mathbf{r}) + \frac{M_{\eta,1}^{(2),+}}{N_\eta^+} \psi_{cv}^{\eta,1}(\mathbf{r}), \tag{10}$$

where $N_\eta^+ = \left| M_{\eta,0}^{(1),+} + M_{\eta,1}^{(2),+} \right|$. The current injection rate at K valley for the two-color $\sigma^+$ polarized beam is

$$\frac{d}{dt} \mathbf{J}_K^{+,+} = e \int \langle \hat{\mathbf{v}} \rangle_{\eta'}^K \delta(E_{cv}^\eta / \hbar - 2\omega) d\eta' = R_X^K(\omega) \mathbf{m}_+, \tag{10}$$

where $\langle \hat{\mathbf{v}} \rangle_\eta^K$ is the expectation value of velocity operator $\hat{\mathbf{v}} = \hat{\mathbf{p}}/m_0 = -i\hbar \nabla_\mathbf{r}/m_0$ and $\langle \hat{\mathbf{v}} \rangle_\eta^K = \langle f_\eta^{+,+} | \hat{\mathbf{v}} | f_\eta^{+,+} \rangle = M_{\eta,0}^{(1),+} \left[ M_{\eta,1}^{(2),+} \right]^* \langle \psi_{cv}^{\eta,1} | \hat{\mathbf{v}} | \psi_{cv}^{\eta,0} \rangle + $ c.c. . The quantum interference of s- and p-state channels excited by one- and two-photon $\sigma^+$ process

respectively, gives a finite expectation value of velocity operator, which gives rise to current. Here $R_X^K = \frac{e\mu_{cv}}{\hbar}|\langle\hat{v}\rangle_\eta^K|$, with the direction of current $\mathbf{m}_+ = \sin(2\varphi_\omega - \varphi_{2\omega})\hat{\mathbf{x}} + \cos(2\varphi_\omega - \varphi_{2\omega})\hat{\mathbf{y}}$.

The current injection rate at $-K$ valley for two $\sigma^+$ beams can also be obtained in similar analysis as illustrated in **Fig. 4(b)**, with $\frac{d}{dt}\mathbf{J}_{-K}^{+,+} = R_X^{-K}(\omega)\mathbf{m}_-$ where $\mathbf{m}_- = \sin(2\varphi_\omega - \varphi_{2\omega})\hat{\mathbf{x}} - \cos(2\varphi_\omega - \varphi_{2\omega})\hat{\mathbf{y}}$. The valley-polarized current injection rate is $\frac{d}{dt}\mathbf{J}_V^{+,+} = \frac{d}{dt}(\mathbf{J}_K^{+,+} - \mathbf{J}_{-K}^{+,+})$. In **Fig. 4 (c)** the current injection rate with respect to the excess photon energy $2\hbar\omega - E_g$ in $K/-K$ valley is demonstrated. For such excitation scheme, the current at K valley is much stronger than in $-K$ valley, the valley-polarized current is mainly contributed by the K valley carriers under the two $\sigma^+$ beams. Comparing the injection rate in both interacting (solid curves) and non-interacting case (dash curves), we showed that the electron-hole interaction significantly enhances the injection rate.

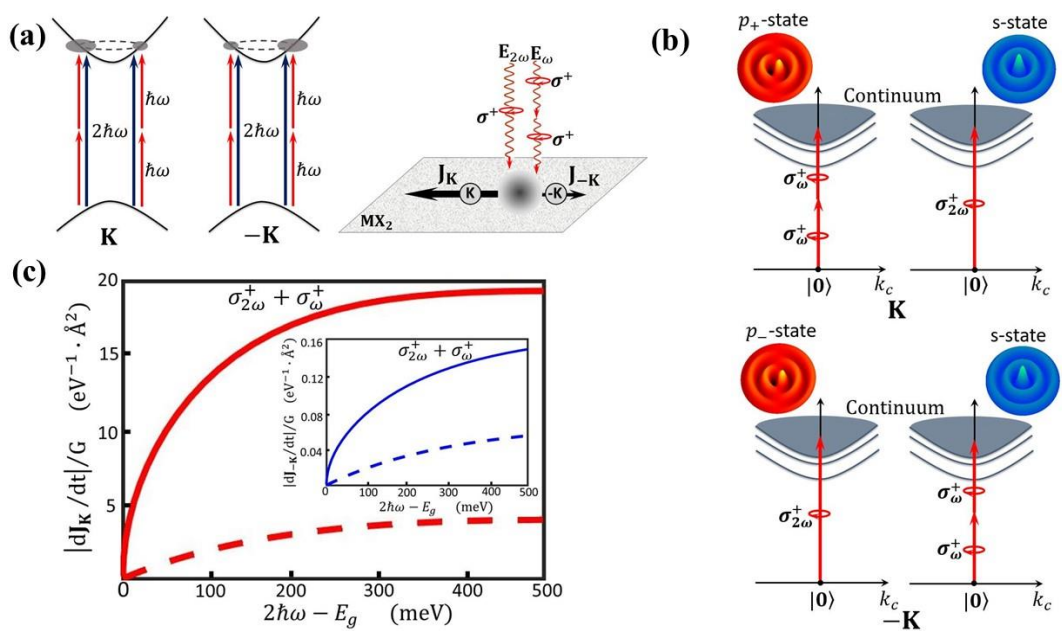

**Figure 4**: **(a)** Current injection by one- and two-photon interference scheme. Left: excitation with two linear polarized beams in non-interacting case. The two beams interference creates imbalance carrier population at $\pm\mathbf{k}$ in a valley. Right: two circularly polarized beams generates different magnitudes of currents from K valley

inducing valley-polarized current. **(b)** For two $\sigma^+$ circularly polarized excitation in interacting case, at K (−K) valley the two-photon beam will excite $p_+$-state ($s$-state) continuum together with one-photon transition to s-state ($p_-$-state). $\sigma^+_{2\omega}$ and $\sigma^+_\omega$ denote the circular polarization for one- and two-photon transitions respectively. **(c)** Current injection rate at K valley from $\sigma^+$ co-circularly two-color field with (without) electron-hole interaction displayed in solid (dash) curves. Inset: current injection rate at −K valley from the same co-circular polarization. Here $G = \frac{e^4}{\hbar^2} E_{2\omega} E_\omega^2$, $p_\pm$-state has angular momentum quantum number $m = \pm 1$.

## 6. Discussions

Although the monolayer TMDs discussed through the paper is naturally a material family with strong electron-hole Coulomb interaction, it is still meaningful to discuss from the non-interacting case, which is due to the following reasons. (1) In the non-interacting case, the discussion of two-photon transition in section 2 allows to reveal the meaningful two-photon polarization reversal picture in **k** space. Such a reversal is microscopically described by the competition between the $\sigma^+$ contribution from conduction and valence band $M_{cv}^+(\mathbf{k})$ and the $\sigma^-$ contribution from remote bands $M_r^-(\mathbf{k})$. While in the interacting case, the two contributions can only be distinguished in the view of rotational symmetry (s or p-state) of electron-hole continuum states. Although in layered TMDs the Coulomb interaction is indeed strong, the microscopic picture to distinguish the two contributions is general, which can be pedagogical and also potentially useful in other systems. Specifically, in the topic optical current injection schemes, most of the available literature discusses it from the **k** space point of view, and the understanding in non-interacting case shall be helpful in studying the current injection scheme involving two-photon process. (2) From non-interacting to interacting case, we tried to provide a pedestrian approach to understand the system, from a relatively straight forward band picture to a more complicated version that attractive Coulomb interaction is included. Such a consideration of arranging the manuscript is also for the completeness of discussions of the material. (3) In experiment design, when performing the nonlinear optical measurements we may put materials with large dielectric constants on both top and bottom of the monolayer TMDC. It can create a sandwiched structure to reduce the effective Coulomb interaction of the TMDC, which resembles the weak interaction limit of the material (see, e.g., ref [72]). Such design provides the possibility to study the system in the weak interacting limit.

As discussed in above sections, both the two-color quantum interference and SHG process involve one- and two-photon transitions from orthogonal angular momentum channels by circularly polarized beams. In **Table II** we summarized the relevant processes in electron-hole continuum regime that are involved. By the table we provide a pedestrian approach to connect the related optical phenomena together as we have discussed through the paper.

|  | Circular Polarization | e-h Continuum States | Remark (K Valley) |
|---|---|---|---|
| One-photon | $\sigma_{2\omega}^+$ | $s$ | $\left|M_{\eta,0}^{(1),+}\right| > \left|M_{\eta,1}^{(1),-}\right|$ |
|  | $\sigma_{2\omega}^-$ | $p$ |  |
| Two-photon | $\sigma_{\omega}^-$ | $p$ | $\left|M_{\eta,1}^{(2),+}\right| \gg \left|M_{\eta,0}^{(2),-}\right|$ |
|  | $\sigma_{\omega}^+$ | $s$ |  |
| SHG | $\sigma_{\omega}^+ + \sigma_{\omega}^+ \to \sigma_{2\omega}^-$ | $p$ | $\left|S_{\eta,1}^{++,-}\right| > \left|S_{\eta,1}^{--,+}\right|$ |
|  | $\sigma_{\omega}^- + \sigma_{\omega}^- \to \sigma_{2\omega}^+$ | $s$ |  |
| Two-color interference | $\sigma_{2\omega}^+ + \sigma_{\omega}^+$ | $s + p$ | valley current generation |

**Table II**: Optical processes of one-photon and two-photon transition, SHG, and two-color quantum interference of electron-hole continuum at K valley. $\sigma_{2\omega}^\pm$ and $\sigma_\omega^\pm$ denote the circular polarization for one- and two-photon transitions. Optical selection rules are given when combined the excitation scheme for circular polarized light at the second column with the continuum states for orthogonal angular momentum channels at third column. See also the discussion in the main text.

## 7. Conclusion

In the paper we studied the two-photon transition and second-harmonic generation process in electron-hole continuum regime in monolayer 2D semiconductors. We demonstrated a two-photon selection rule reversal of circularly polarized light by the competition of two contributions from different bands. By including the electron-hole interaction, we showed that the interaction qualitatively changes selection rules via the two orthogonal angular momentum channels of excitation, with significant enhancement effect. We also discussed the valley-polarized current injection by the one- and two-photon quantum interference with the interaction. Our study provides a coherent and comprehensive view for understanding the relevant nonlinear optical processes in electron-hole continuum regime in monolayer 2D semiconductors.

**Acknowledgments**

The work is supported by the Croucher Foundation (Croucher Innovation Award), the Research Grants Council (HKU17305914P, C7036-17W).

**Author Contributions**

P.G. and W.Y. conceived the project. P.G. performed the calculations, with input from all authors. P.G., H.Y. and W.Y. wrote the manuscript.